\documentclass[sigconf]{acmart}
 
\usepackage{natbib}
\usepackage{multirow}
\usepackage{graphicx}
\usepackage{markdown}
\usepackage{subcaption}
\usepackage{multicol}

\usepackage{todonotes}

\definecolor{kentuckyblue}{RGB}{0, 93, 170}
\definecolor{canada_red}{RGB}{216, 6, 33}

\setcopyright{acmlicensed}
\copyrightyear{2025}
\acmYear{2025}
\setcopyright{rightsretained}
\acmConference[FAccTRec@RecSys’25]{FAccTRec@RecSys’25 Workshop}{September 16-19 2025}{Prague}
\acmDOI{}
\acmISBN{}

\begin{document}
\title{Envy-Free but Still Unfair: Envy-Freeness Upto One Item (EF-1) in Personalized Recommendation}

\author{Amanda Aird}
\email{amanda.aird@colorado.edu}
\affiliation{%
  \institution{Department of Information Science, University of Colorado, Boulder}
  \city{Boulder}
  \state{Colorado}
  \country{USA}
  \postcode{80309}
}

\author{Ben Armstrong}
\email{barmstrong@tulane.edu}
\affiliation{%
  \institution{Department of Computer Science, Tulane University}
  \city{New Orleans}
  \state{Louisiana}
  \country{USA}
  \postcode{70118}
}

\author{Nicholas Mattei}
\email{nsmattei@tulane.edu}
\affiliation{%
  \institution{Department of Computer Science, Tulane University}
  \city{New Orleans}
  \state{Louisiana}
  \country{USA}
  \postcode{70118}
}

\author{Robin Burke}
\email{robin.burke@colorado.edu}
\orcid{0000-0001-5766-6434}
\affiliation{%
  \institution{Department of Information Science, University of Colorado, Boulder}
  \city{Boulder}
  \state{Colorado}
  \country{USA}
  \postcode{80309}
}

\begin{abstract}
    Envy-freeness and the relaxation to Envy-freeness up to one item (EF-1) have been used as fairness concepts in the economics, game theory, and social choice literatures since the 1960s, and have recently gained popularity within the recommendation systems communities. In this short position paper we will give an overview of envy-freeness and its use in economics and recommendation systems; and illustrate why envy is not appropriate to measure fairness for use in settings where personalization plays a role. 
\end{abstract}

\maketitle

\section{Introduction}

Envy is part of the human experience: that feeling you might have gotten when your sibling got a candy bar growing up and you did not. The idea of desiring something given to or possessed by others is formalized in distributive economics as the notion of an \emph{envy} with respect to allocations \cite{foley1966resource} of a set of resources. In the economics and social choice literature, the absence of envy (\textit{envy-freeness}) has been one of several metrics used to decide whether or not an allocation of goods (land, resources) is \emph{fair} or acceptable \cite{Thomson:Allocation}. This concept has recently been imported into the recommendation systems space as a fairness metric to judge whether a set of items recommended to a user (their allocation) is fair. In this short position paper we will give an overview of envy-freeness and its use in economics and recommendation systems; and illustrate why envy is not appropriate to measure fairness for use in settings where personalization plays a role.

\section{Measuring Envy}

Envy-freeness (EF) is a concept from the economics of fair division which measures how satisfied agents are with an allocation. It was first introduced by \citet{foley1966resource} as a means to measure allocations of land and other public goods. The metric originates in the idea of distributive justice \cite{arnsperger1994envy}: when dividing up resources which may have differing value to different agents no agent should prefer (envy) the resources given to another agent over those they received. This is formalized as the \emph{envy-free} condition, which holds when no agent is envious of any other relative to a given allocation.

Informally, envy-freeness is a property of an allocation. Imagine we need to divide up some candy between three agents. We allocate some pieces to each, and if every agent would rather retain their set of candy than trade it with another, the allocation is said to be envy-free. Likewise, the relaxation of envy-freeness upto one item (EF-1) is the same property, but where we can remove an item from another agent's set of candy (typically the most valuable item). Envy-freeness is an intiutivly appealing concept as it only asks an agent to evaluate their bundle against others according to their own utility, so we do not need to make inter-agent comparisons of utility \cite{aziz2019constrained}.

The classical model studies a finite and small set of items $X = \{x_1, \ldots, x_m\}$ to be distributed among a small set of agents $A = \{a_1, \ldots, a_n\}$ \cite{arnsperger1994envy,amanatidis2023fair}.
Each agent is able to evaluate the \emph{utility} of each item; we have $U_{a_i}(x_i)$ and $U_{a_i}(\{x_i, x_j, \ldots \})$ as, respectively, agent $a_i$'s value for a single item $x_i$, and a bundle of items $\{x_i, x_j, \ldots \}$. Because each agent values bundles independently, there is no need to compare valuations across agents. Most literature assumes that utilities are \textit{positive}, i.e. an agent prefers to have any item rather than not have it, and \textit{additive}, i.e., more items are better \cite{Thomson:Allocation}. Current research in the area typically tries to relax these assumptions in various ways and come up with strategies or algorithms for allocating these resources under various constraints on fairness, efficiency, endowments, etc. \cite{amanatidis2023fair}.

\begin{figure*}[h!]
    \centering
    \includegraphics[width=\linewidth]{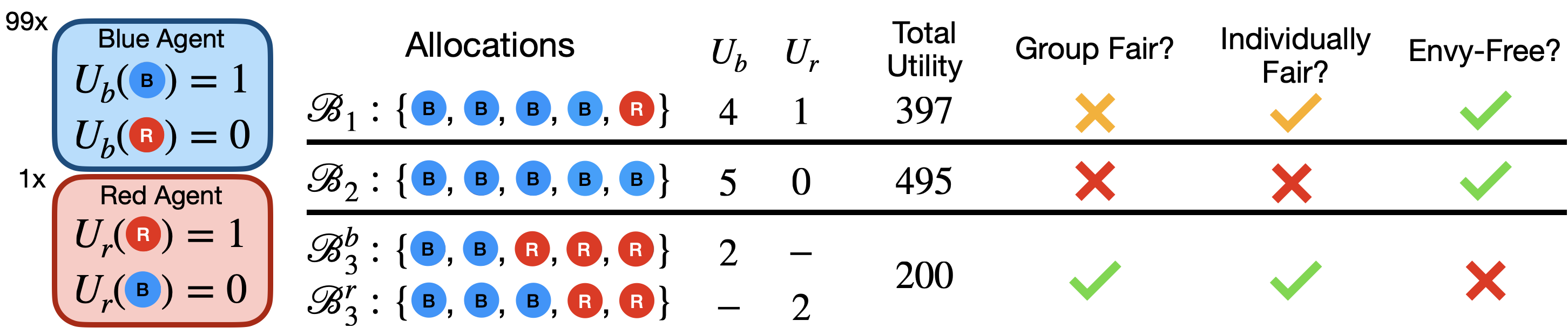}
    \caption{Three sets of recommendation bundles and utilities for each agent type. The examples demonstrate that there is no inherent relationship between fairness, utility, and envy.}
    \label{fig:example}
\end{figure*}

Formally, an an allocation of all items $\mathcal{A}$, where $\mathcal{A}_{a_i}$ is the allocation of items to $a_i$, is \emph{envy-free} if and only if there exists no agent $a_i$ such that $\forall a_j \in (A \setminus \{a_i\}): U_{a_i}(\mathcal{A}_{a_j}) > U_{a_i}(\mathcal{A}_{a_i})$. That is, an allocation is envy-free if every agent receives as much utility from the bundle of items they have received as they would from the allocation given to any other agent. We can quantify envy by calculating the number of agents for whom the envy constraint is not met. Note that we do not add up the amount of envious utility, one of the key aspects of envy is it only requires one agent to compare their utilities for different potential bundles, nicely sidestepping the common problem of inter-agent utility comparisons \cite{foley1966resource}.

For indivisible goods, the existence of an envy-free allocation is not guaranteed and even when one exists, calculating it may be intractable. As a result, several relaxations of EF have been introduced, including the popular \textit{envy-freeness up to one good} (EF1): An allocation satisfies EF1 if, for any agent $a_i$ that envies $a_j$ there exists an item in $\mathcal{A}_{a_j}$ such that removing it from $a_j$'s allocation will remove $a_i$'s envy of $a_j$. An EF1 allocation is known to (1) always exist, and (2) be computable in polynomial time \cite{budish2011combinatorial}. 

However, there are several shortcomings with relying on the no-envy condition alone. For instance, the empty allocation, i.e., one where no agent receives any allocation, is envy-free since all agents have both the same utility for their bundle and all other agent bundles --- namely, 0 \cite{Thomson:Allocation}. Likewise, envy as a concept is built on the idea that items have (at least somewhat) comparable values between agents. Individuals may value one type of candy bar differently from one another, but (likely) have a similar order of magnitude to the candy bar valuations of other agents. Finally, it assumes that these sets are small and that there is some competition. If there is no contest between items, e.g., each agent likes a completely different type of music than all others, then envy-free allocations are relatively easy to find. Within the economics literature on envy-freeness, the distinction between envy and fairness has long been known. Early work has shown that, while both useful, these two concepts are easily confused yet fundamentally distinct \cite{holcombe1997absence}.

\section{Envy in Recommender Systems}

Recent work in recommender systems has adapted envy-freeness to recommendation settings in several different ways. For example, there is research in group recommendation in which packages of items are recommended to a group of users and the question is the envy among members of a group \cite{serbos2017fairness}. This setting is rivalrous in the sense that all group members share the recommended bundle of items, e.g., travel destinations to be experienced together.

Other works in the recommender systems space include \citet{li2024feir} and \citet{dickerson2014computational} which look at recommendation in ride-sharing, job-matching, organ-matching, and other two-sided markets. Envy is more salient in these settings as the allocations themselves are more rivalrous and intrinsic values are easier to determine \cite{li2024feir}. Likewise, \citet{saito2022fair} explicitly use envy-freeness as a measure of exposure in rankings. This is much more in the spirit of envy from the economics literature as it is allocating a scarce finite resource, namely exposure.

A less convincing line of work is one that uses EF1 as a measure of consumer-side fairness. The \emph{FairRec} systems and related work \cite{patro2020fairrec,biswas2021toward} combines provider-side fairness while using consumer-side fairness defined in terms of EF1 as a guarantee that fairness for providers does not have a significant impact on fairness for consumers. In similar work, \citet{do2022online} highlight that providing parity of exposure for providers does not conflict with giving envy-free recommendations, but that more personalized provider-side fairness metrics such as equity of exposure do not admit envy-free outcomes. They show that in certain settings, nearly envy-free recommendations can be provided and verified, while minimizing the amount of user preference information required \cite{do2022online}.

\subsection{Where Envy-Freeness is Not Enough}

Despite these lines of research, it is not difficult to show that envy runs into trouble capturing basic notions of fairness in recommendation. Consider the following small example of the recommendations delivered in a stylized recommender system. 
\begin{description}
    \item[Setting:] Assume two types of items: blue (B) and red (R), and two types of users: ``blue'' users ($a_b$) with $U_{a_b}(B) = 1$ and $U_{a_b}(R) = 0$, and ``red'' users ($a_r)$ with $U_{a_b}(B) = 0$ and $U_{a_b}(R) = 1$. There are 99 blue users and 1 red user where red users are considered to be a protected group. Each user receives the same bundle of recommendations. 
    \item[Bundle 1:] $\mathcal{B}_1 = \{B, B, B, B, R\}$; 4 blue items and 1 red item. Blue agents receive a total utility of $U_{a_b}(\mathcal{B}_1) = 4$, while the red agent receives $U_{a_r}(\mathcal{B}_1) = 1$. The fraction of relevant recommendations from Bundle 1 is $0.99*\frac{4}{5} + 0.01*\frac{1}{5} = 0.794$, the total utility summed across all items and agents is $397$, and no agent is envious.
    \item[Bundle 2:] $\mathcal{B}_2 = \{B, B, B, B, B\}$; 5 blue items. Now we have $U_{a_b}(\mathcal{B}_2) = 5$ and $U_{a_r}(\mathcal{B}_2) = 0$. The fraction of relevant recommendations from Bundle 2 is $0.99*\frac{5}{5} + 0.01*\frac{0}{5} = 0.99$, the total utility is $495$, and no agent has envy.
\end{description}

This setup is depicted in \autoref{fig:example}. 
Bundle 2 is dominant because it provides more agents with relevant recommendations and, hence, higher total utility while being envy-free like Bundle 1.

However, this determination is at odds with what group fairness requires. Recall that the red users are a protected group, clearly a minority in this example. A typical group fairness metric compares protected and unprotected groups based on the difference in their accuracies \cite{ekstrand2022fairness}. For Bundle 1, that difference is $\frac{3}{5}$. For Bundle 2, that difference is 1, making it the least fair option. The fact that the red user has no envy relative to Bundles 1 and 2 is not because of fair treatment, but because their tastes differ from the majority.

Consider individual fairness on the consumer side. A common metric is the Gini coefficient of accuracies, measuring whether the benefits of the system are distributed evenly across the user population \cite{ekstrand2022fairness}; low Gini values are fairer. Bundle 1 achieves a Gini of 0.0075; Bundle 2, a value of 0.01. So, Bundle 2 is also less individually fair, which makes sense because difference between the worst off and best off users is greater.

The example illustrates that envy alone is not a sufficient metric to capture consumer-side fairness, whether considered at the group or the individual level. Now consider an additional case in which the two groups of users get different recommendations. Two blue items delivered to each blue user $\mathcal{B}_{3}^{b} = \{B, B, R, R, R\}$, and two red items delivered to the red user $\mathcal{B}_{3}^{r} = \{R, R, B, B, B\}$. Now we have $U_{a_b}(\mathcal{B}_{3}^{b}) = 2$ and $U_{a_r}(\mathcal{B}_3^r) = 2$. The fraction of relevant recommendations from Bundle 3 is $0.99*\frac{2}{5} + 0.01*\frac{2}{5} = 0.4$, the total utility is $200$, but envy is 100\%. Every blue agent would prefer the red agent's bundle and the red agent would prefer any blue agent's bundle. Still, we have perfect fairness, every agent has the same utility, so both group and individual fairness are maximized. 

For provider-side fairness, let us  invert the sense of our example and consider two types of consumers arriving at a recommendation platform: blue and red. We have blue providers, whose products are of interest to the blue users, and red providers, appealing to the red users. There is no utility for a blue provider in having their product recommended to a red user because no purchase would ever result. Our different bundles amount to different distributions of user characteristics. The provider-fairness question becomes isomorphic to our prior consumer-side discussion. Envy among providers is not necessarily correlated with fairness, either individual or group.

\section{Conclusion}

Thus, we conclude that envy and its cousin EF1 are not, in general, appropriate metrics for fairness in recommendation. In our example, we see that a set of recommendations may be high in envy and fair or low in envy and still unfair. This conclusion is strongest in the realm of consumer-side fairness because the assumption of personalized and diverse consumer utility is definitional for recommender systems, as a technology of personalization. However, as we show, diverse utilities relative to recommendation opportunities may exist for providers as well, and in such cases, envy fails to correlate with fairness on the provider side. Understanding how to mix competing (and possibly complimentary) notions of fairness in recommender systems is an important direction that we are actively working on by incorporating more notions from social choice along with stakeholder interactions \cite{farastu2022pays, aird2024dynamic, smith2023many}.

\begin{acks}
Authors Burke and Aird were supported by the National Science Foundation under grant awards IIS-1911025 and IIS-2107577. Nicholas Mattei and Ben Armstrong were supported by NSF Grant IIS-III-2107505 and in part by NSF Awards IIS-RI-2134857, IIS-RI-2339880 and CNS-SCC-2427237 as well as the Harold L. and Heather E. Jurist Center of Excellence for Artificial Intelligence at Tulane University and the Tulane University Center for Community-Engaged Artificial Intelligence  
\end{acks}
\bibliographystyle{ACM-Reference-Format}
\bibliography{ef.bib}
\end{document}